\documentclass[aps,pre,twocolumn,superscriptaddress,showpacs]{revtex4}
\usepackage{graphicx}
\begin{document}

\title{Spontaneous symmetry breaking of solitons trapped in a double-channel
potential}
\author{M.~Matuszewski}
\affiliation{Institute of Theoretical Physics, Physics Department, Warsaw University, Ho%
\.{z}a 69, PL-00-681 Warsaw, Poland}
\author{B.~A.~Malomed}
\affiliation{Department of Interdisciplinary Sciences, School of Electrical Engineering,
Faculty of Engineering, Tel Aviv University, Tel Aviv 69978, Israel}
\author{M.~Trippenbach}
\affiliation{Institute of Theoretical Physics, Physics Department, Warsaw University, Ho%
\.{z}a 69, PL-00-681 Warsaw, Poland}
\affiliation{Soltan Institute for Nuclear Studies, Ho%
\.{z}a 69, PL-00-681 Warsaw, Poland}

\begin{abstract}
We consider a two-dimensional (2D) nonlinear Schr\"{o}dinger equation with
self-focusing nonlinearity and a quasi-1D double-channel potential, i.e., a
straightforward 2D extension of the well-known double-well potential. The
model may be realized in terms of nonlinear optics and Bose-Einstein
condensates. The variational approximation (VA) predicts a bifurcation
breaking the symmetry of 2D solitons trapped in the double channel, the
bifurcation being of the \textit{subcritical} type. The predictions of the
VA are confirmed by numerical simulations. The work presents the first
example of the spontaneous symmetry breaking (SSB) of 2D solitons in any
dual-core system.
\end{abstract}

\maketitle

\section{Introduction}

The creation of Bose--Einstein condensates (BECs) in vapors of alkali metals
has opened up a unique opportunity to investigate nonlinear interactions of
atomic matter waves. In particular, an important physical problem is to
develop methods allowing one to create and control matter-wave solitons in
the experiment. One-dimensional (1D) dark \cite{1fsol}, bright \cite{2bright}%
, and gap-mode \cite{ober} solitons have already been created.

Another fascinating phenomenon, viz., the Josephson effect in BEC,
was observed in condensates loaded in a double-well potential.
Tunneling through a potential barrier usually occurs in quantum
systems on a nanoscopic scale, while the Josephson effect features
the tunneling of \emph{macroscopic} wave functions describing
intrinsically coherent states \cite{JJ}. This phenomenon has been
observed in sundry systems, such as a pair of superconductors
separated by a thin insulator (the Josephson effect proper)
\cite{JJexp}, and two reservoirs of superfluid helium connected by
nanoscopic apertures \cite{JJ1,JJ2}. Recently, the first
successful implementation of a bosonic Josephson junction formed
by two weakly coupled BECs in a macroscopic double-well potential
was reported \cite{JJober}. In contrast to hitherto realized
Josephson systems in superconductors and superfluids, interactions
between tunneling particles plays a crucial role in the junction
implemented in the BEC setting, the effective nonlinearity induced
by the interactions giving rise to new dynamical regimes in the
tunneling. In particular, anharmonic Josephson oscillations were
predicted \cite{JJ3,JJ4,JJ5}, provided that the initial population
imbalance in the two potential wells falls below a critical value.
The dynamics change drastically for the initial population
difference exceeding the threshold of the macroscopic quantum
self-trapping and thus inhibiting large-amplitude Josephson
oscillations \cite{JJ6,JJ7,JJ8}. These two different regimes have
been experimentally demonstrated in the BEC-based
Josephson-junction arrays \cite{JJ9,JJ10,JJ11}.

This dynamics can be well explained by means of a simple model
derived from the Gross-Pitaevskii equation ({GPE)}. Two equations
for the self-interacting BEC amplitudes, linearly coupled by
tunneling terms, describe the dynamics in terms of the inter-well
phase difference and population imbalance. As mentioned above, the
nonlinearity specific to BEC gives rise to the ``macroscopic
quantum self-trapping"
effect, in the form of a self-maintained population imbalance \cite%
{JJ7,JJT1,JJT2,JJT3}. In order to derive a reduced two-state model, one
needs to find eigenstates of the corresponding GPE and perform stability
analysis for them. Such analysis can be readily performed for the
square-shaped double-well potential, where analytic solutions are available
\cite{DSW,JJT2,JJT3}. In this case, the point of the symmetry-breaking
bifurcation, where asymmetric solutions emerge, can be found exactly.

The present paper addresses the symmetry-breaking bifurcation and the
existence and stability of asymmetric states in a two-dimensional (2D)
system, which is a direct extension of the familiar double-well model \cite%
{JJ7,JJ8}. The corresponding potential is shown in Fig. \ref{fig0} below. It
features the double-square-well shape in the $x$ direction, being uniform
along $y$. To our knowledge, the spontaneous symmetry breaking (SSB) has
never been studied in 2D systems before. By means of the variational
approximation (VA), we will find regions where stable asymmetric solitons
exist, and the prediction will be then confirmed by direct simulations.

The paper is organized as follows. The model in introduced in Sec. II, where
its physical interpretations are outlined too, in terms of BEC and nonlinear
optics. Then, in Sec. III, we derive variational equations and analyze their
solutions, which predict the SSB of a \textit{subcritical }type. At the end
of that Section, we compare the result with those for the CW
(continuous-wave, i.e., as a matter of fact, one-dimensional) states, for
which the SSB bifurcation is of a different type, being \emph{supercritical}%
. In Sec. IV, we compare predictions of the VA with numerical results, and
Sec. V concludes the paper.

\section{The model}

\begin{figure}[tbp]
\includegraphics[width=7.5cm]{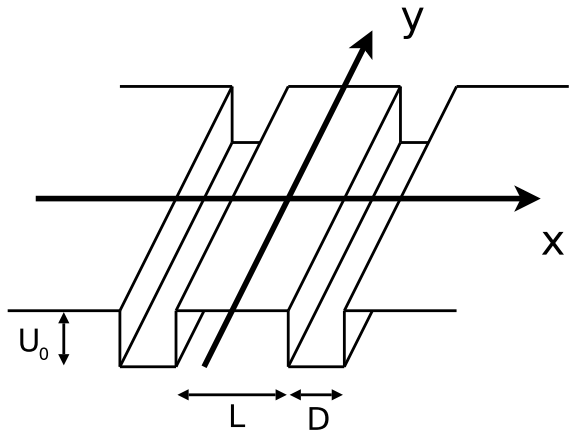}
\caption{The shape of the quasi-one-dimensional double-well potential, $%
U(x,y)$.}
\label{fig0}
\end{figure}

The starting point is the 2D equation in a normalized form, with the
self-attracting cubic nonlinearity,
\begin{equation}
i\Psi _{t}+\frac{1}{2}\left( \Psi _{xx}+\Psi _{yy}\right) -U(x)\Psi +|\Psi
|^{2}\Psi =0,  \label{eq1}
\end{equation}%
where the quasi-1D double-well potential is taken as
\begin{equation}
U(x)=\left\{
\begin{array}{ll}
0, & |x|~<\frac{1}{2}L~\mathrm{and~}|x|~>\frac{1}{2}L+D, \\
-U_{0}, & \frac{1}{2}L<|x|~<\frac{1}{2}L+D,%
\end{array}%
\right.  \label{DC}
\end{equation}%
with $D$, $U_{0}$ and $L$ being, respectively, the width and depth of each
well, and the width of the barrier between them, see Fig. \ref{fig0}.

Equation (\ref{eq1}) admits three different physical interpretations. In
terms of BEC, it is the Gross-Pitaevskii equation (GPE), for a
pancake-shaped (nearly flat) self-attractive condensate (in fact, this
implies $^{7}$Li , in which the scattering length can be readily made
slightly negative by means of the Feshbach resonance \cite{2bright}), with
potential (\ref{DC}) acting in the $\left( x,y\right) $ plane. In nonlinear
optics, the evolutional variable, $t$, is actually the propagation distance
(usually denoted by $z$). Then, one possible interpretation is that Eq. (\ref%
{eq1}) is the nonlinear Schr\"{o}dinger equation (NLSE) which, adopting the
ordinary paraxial approximation for the diffraction, governs the stationary
transmission of an optical signal in the spatial domain (i.e., in the bulk
medium) with the self-focusing Kerr nonlinearity; then, the double-channel
potential corresponds to two waveguiding slabs embedded in the 3D medium. An
alternative optical interpretation is valid in the temporal domain, where
Eq. (\ref{eq1}) describes the light propagation in a nonlinear planar
waveguide with a pair of embedded guiding channels. In the latter case, $x$
is the transverse coordinate, while $y$ is actually the reduced time, $t-z/%
\bar{V}$ ($t$ is physical time, and $\bar{V}$ is the mean group velocity of
the carrier wave), assuming that the group-velocity dispersion in the planar
waveguide is anomalous.

According to the above interpretations, possible localized solutions of
equation (\ref{eq1}), if it is considered as the GPE, will be interpreted as
matter-wave solitons. If the equation is realized as the NLSE in optical
media, the localized solutions will be either spatial or spatiotemporal
solitons (in the bulk and planar waveguide, respectively; a review of the
topic of multidimensional optical solitons can be found in Ref. \cite{review}%
).

We are looking for stationary solutions as $\Psi (x,y,t)=e^{-i\mu t}\Phi
(x,y)$, where real function $\Phi (x,y)$ satisfies equation
\begin{equation}
\mu \Phi +\frac{1}{2}\left( \Phi _{xx}+\Phi _{yy}\right) -U(x)\Phi +\Phi
^{3}=0,  \label{Phi}
\end{equation}%
which can be derived from the Lagrangian,
\begin{eqnarray}
L_{\mathrm{stat}} &=&\frac{1}{2}\int \int dxdy\left[ \mu \Phi ^{2}-\frac{1}{2%
}\left( \Phi _{x}^{2}+\Phi _{y}^{2}\right) -\right.  \nonumber \\
&&\left. -U(x)\Phi ^{2}+\frac{1}{2}\Phi ^{4}\right] .  \label{Lstat}
\end{eqnarray}%
Our variational approximation (VA) will be based on this Lagrangian.

\section{Variational approximation}

\subsection{Symmetry breaking of the two-dimensional solitons}

To apply the VA (a detailed account of the technique can be found in review
\cite{Progress}), we assume a situation with two very narrow and deep
channels, separated by a broad barrier, i.e., $D\ll L$ in Eq. (\ref{DC}).
The \textit{ansatz} describing the soliton field configuration consists of
two parts. First, inside each channel, i.e., in regions $\left\vert x\mp
\left( L+D\right) /2\right\vert <D/2$, we adopt the trial function
\begin{equation}
\Phi (x,y)=A_{\pm }\cos \left( \pi \frac{x\mp \left( L+D\right) /2}{D}%
\right) \exp \left( -\frac{y^{2}}{2W^{2}}\right) ,  \label{inner}
\end{equation}%
where $A_{\pm }$ and $W$ are three real variational parameters. In this
expression, we assume that the wave function has different amplitudes but
equal longitudinal widths, $W$, in both channels. In direction $x$, ansatz (%
\ref{inner}) emulates the ground state of a quantum-mechanical particle in
an infinitely deep potential box, therefore it vanishes at edges of the
channel. In direction $y$, the ansatz is assumed to be a self-trapped
soliton, approximated by the Gaussian. The form of the ansatz outside the
channels is also suggested by quantum mechanics, emulating a superposition
of exponentially decaying ground-state wave functions behind the edges of
deep potential boxes,
\begin{equation}
\Phi (x,y)=\sum_{+,-}A_{\pm }\exp \left( -\sqrt{-2\mu }\left\vert x\mp \frac{%
L+D}{2}\right\vert -\frac{y^{2}}{2W^{2}}\right) ,  \label{outer}
\end{equation}%
where amplitudes $A_{\pm }$ and width $W$ are the same as in Eq. (\ref{inner}%
).

The substitution of the inner and outer parts of the ansatz, Eqs. (\ref%
{inner}) and (\ref{outer}), into Lagrangian (\ref{Lstat}), and subsequent
integration over $x$ and $y$ lead to the following effective Lagrangian:%
\[
\frac{2}{D\sqrt{\pi }}L_{\mathrm{eff}}=
\]%
\begin{eqnarray}
&&\sum_{+,-}\left( \frac{\mu +U_{0}}{2}A_{\pm }^{2}W-\frac{A_{\pm }^{2}}{8W}+%
\frac{3}{16\sqrt{2}}A_{\pm }^{4}W\right)  \nonumber \\
&&+\frac{4\sqrt{-2\mu }}{D}e^{-\sqrt{-2\mu }\left( L+D\right) }A_{+}A_{-}W.
\label{L_eff}
\end{eqnarray}%
Here, we have adopted the Thomas-Fermi approximation in the $x$ direction
(but not along $y$), by omitting the corresponding kinetic-energy term, $%
-(1/2)\Phi _{x}^{2}$, in the Lagrangian density.

Note that ansatz (\ref{inner}) includes the cosine trial function
with a constant width. This assumption is relevant for a deeply
bound quantum state (as mentioned above, the ansatz was modeled on
the pattern of the ground state in the infinite deep box), but not
when the energy eigenvalue $|\mu |$ is small. Indeed, if one tries
to predict, by means of this ansatz, a bound state of a particle
in a finite-depth rectangular potential well in ordinary (linear)
quantum mechanics, one would arrive at a conclusion that the bound
state appears only starting from a minimum finite value of
$U_{0}$, which is, as a matter of fact, the kinetic energy in the
$x$ direction. A commonly known exact result is that there is no
threshold for the existence of a bound state in any symmetric
potential well, even if it is arbitrarily shallow. However, this
unphysical effect disappears in the Thomas-Fermi approximation,
which was adopted above.

To simplify the notation, we introduce new parameters:
\begin{equation}
\epsilon \equiv \mu +U_{0}  \label{eps}
\end{equation}%
(notice that $\epsilon $ may be both positive and negative),
\begin{equation}
\lambda \equiv \left( 2/D\right) \sqrt{-2\mu }\exp \left( -\sqrt{-2\mu }%
\left( L+D\right) \right) ,  \label{lambda}
\end{equation}%
and $N_{\pm }\equiv \left( 3/4\sqrt{2}\right) A_{\pm }^{2}W$. Additionally,
we define%
\begin{equation}
N\equiv \frac{N_{+}+N_{-}}{4\sqrt{\lambda }},~\nu \equiv \frac{N_{+}-N_{-}}{4%
\sqrt{\lambda }}.  \label{nuN}
\end{equation}%
Norms of the wave function (which are proportional to the numbers of trapped
atoms) in the two channels are, according to ansatz (\ref{inner}),
\[
\left\vert \int_{-\infty }^{+\infty }dy\int_{\pm L/2}^{\pm \left(
D+L/2\right) }dx\left( \Phi (x,y)\right) ^{2}\right\vert =\left( \sqrt{\pi }%
/2\right) A_{\pm }^{2}DW,
\]%
hence parameters $N_{\pm }$ are proportional to the populations in the
channels, while $\nu $ determines the population imbalance\textit{.} In
terms of this notation, effective Lagrangian (\ref{L_eff}) transforms into%
\[
\frac{3}{8\sqrt{2\pi \lambda}D}L_{\mathrm{eff}}=
\]%
\begin{eqnarray}
&&\frac{1}{4\sqrt{\lambda }}\sum_{+,-}\left( \frac{\epsilon N_{\pm }}{2}-%
\frac{N_{\pm }}{8W^{2}}+\frac{N_{\pm }^{2}}{4W}\right) +\frac{\sqrt{\lambda
N_{+}N_{-}}}{2}  \nonumber \\
&\equiv &\frac{\epsilon N}{2}-\frac{N}{8W^{2}}+\frac{\sqrt{\lambda }}{2}%
\frac{N^{2}+\nu ^{2}}{W}+\lambda \sqrt{N^{2}-v^{2}}.  \label{Nnu}
\end{eqnarray}

This Lagrangian gives rise to variational equations $\partial L/\partial W=$
$\partial L/\partial N=\partial L/\partial \nu =0$, i.e., respectively,
\begin{eqnarray}
\frac{N}{2\sqrt{\lambda }\left( N^{2}+\nu ^{2}\right) } &=&W,  \label{W} \\
\frac{1}{2}\epsilon -\frac{1}{8W^{2}}+\frac{\sqrt{\lambda }N}{W}+\frac{%
\lambda N}{\sqrt{N^{2}-\nu ^{2}}} &=&0,  \label{N} \\
\nu \left( \frac{\sqrt{\lambda }}{W}-\frac{\lambda }{\sqrt{N^{2}-\nu ^{2}}}%
\right) &=&0.  \label{nu}
\end{eqnarray}%
Equation (\ref{nu}) has two solutions: $\nu =0$, which, according to Eq. (%
\ref{nuN}) corresponds to symmetric solitons, and asymmetric ones, with $\nu
\neq 0$.

In Eqs. (\ref{W}) - (\ref{nu}), it is relevant to consider $N$ (which is
proportional to the total number of atoms) as a given parameter, and $%
\epsilon $ [which is, as the matter of fact, the chemical potential, see Eq.
(\ref{eps})] as an unknown. In this manner, for each value of $N$ we can
find the corresponding vales of $\lambda ,\epsilon ,\nu $ and $W$ [system (%
\ref{W}) - (\ref{nu}) contains only three equations, but, as seen from
definitions (\ref{eps}) and (\ref{lambda}), $\epsilon $ and $\lambda $ are
not independent].

Being interested in asymmetric solutions, with $\nu \neq 0$, and
substituting expression (\ref{W}) for $W$ in Eq. (\ref{nu}), we arrive at an
equation which determines the population imbalance, $\nu $, as a\ function
of $N$:
\begin{equation}
2\sqrt{N^{2}-\nu ^{2}}\left( N^{2}+\nu ^{2}\right) =N.  \label{nu(N)}
\end{equation}%
Notice that this cubic equation for $\nu ^{2}$ does not contain $\lambda ,W$
or $\epsilon $. The most essential issue is when the \textit{spontaneous
symmetry breaking} (SSB) occurs, i.e., at what value of $N$ a nontrivial
solution for $\nu $ appears in Eq. (\ref{nu(N)}).

\begin{figure}[tbp]
\includegraphics[width=8.5cm]{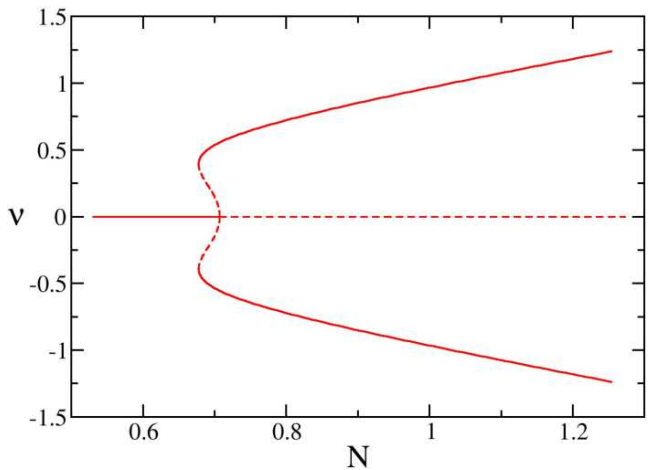}
\caption{(Color online). The subcritical symmetry-breaking bifurcation in
the double-channel model, as predicted by the variational approximation
through Eq. (\protect\ref{nu(N)}). Unstable branches of the solutions are
shown by dashed curves.}
\label{fig1}
\end{figure}

Straightforward analysis of Eq. (\ref{nu(N)}) demonstrates that, with the
increase of $N$, a pair of physical (real) solutions for $\nu $ appears
through a \textit{tangent} (alias \textit{saddle-center)} bifurcation at $%
N=N_{\mathrm{\min }}=(1/2)\sqrt{3\sqrt{3/2}}\approx \allowbreak 0.958$. At a
slightly larger critical value, $N_{\max }=1$, a \textit{subcritical
pitchfork} bifurcation takes place, giving rise to solutions splitting off
from $\nu =0$. The entire bifurcation diagram for Eq. (\ref{nu(N)}) is
displayed in Fig. \ref{fig1}. According to general principles of the
bifurcation theory \cite{JosephIooss}, the picture implies that the
symmetric solution, $\nu =0$, is stable in interval $0<N<1$, and unstable
for $N>1$. Simultaneously, branches of the asymmetric solutions (those with $%
\nu \neq 0$) are unstable as long as they go backward, and become stable
after they turn forward at point $N=N_{\min }$.

This subcritical bifurcation is qualitatively similar to that explored
earlier the model of dual-core nonlinear optical fibers, which is based on a
pair of linearly coupled one-dimensional NLSEs for amplitudes of the
electromagnetic waves in the two cores \cite{Sydney,Progress}. On the other
hand, in an apparently more complex model of parallel-coupled fiber Bragg
gratings, that amounts to a system of four equations with linear and
nonlinear couplings, the SSB is simpler, being \textit{supercritical} (the
emerging branches of asymmetric solutions immediately go forward and are
stable everywhere) \cite{Mak}.

\subsection{Comparison with the one-dimensional (continuous-wave) case}

\begin{figure}[tbp]
\includegraphics[width=8.5cm]{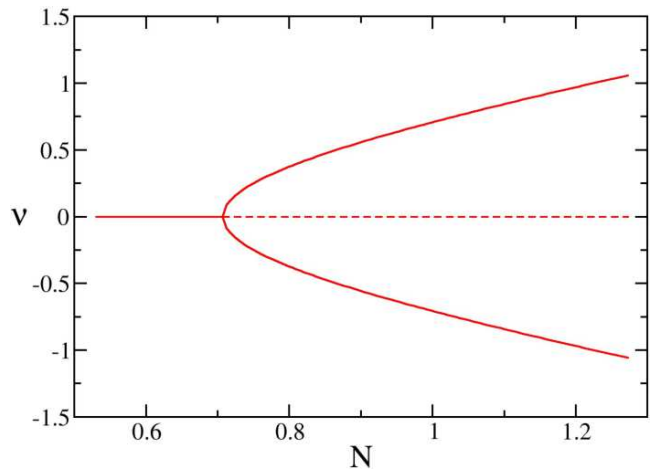}
\caption{(Color online). The supercritical symmetry-breaking bifurcation for
the continuous-wave (CW) states, in the 1D model.}
\label{fig2}
\end{figure}

It is relevant to compare the above results for the SSB of solitons in the
2D model to what can be predicted in the 1D counterpart of the model by the
same type of the VA. The latter corresponds to the ansatz based on Eqs. (\ref%
{inner}) and (\ref{outer}), but with $W=\infty $ (in other words, this is a
CW state in terms of the 2D model). Then, Lagrangian (\ref{L_eff}) reduces
to
\begin{equation}
L_{\mathrm{eff}}=\mathrm{const}\cdot \sum_{+,-}\left( \frac{1}{2}\epsilon
A_{\pm }^{2}+\frac{3}{16\sqrt{2}}A_{\pm }^{4}+2\lambda A_{+}A_{-}\right) .
\label{L1D}
\end{equation}%
Straightforward manipulations with the variational equations generated by
this Lagrangian, $\partial L_{\mathrm{eff}}/\partial A_{+}=\partial L_{%
\mathrm{eff}}/\partial A_{-}=0$, yield a final relation for asymmetric
states:
\begin{equation}
\left( A_{+}^{2}-A_{-}^{2}\right) ^{2}=\left( A_{+}^{2}+A_{-}^{2}\right)
^{2}-2\left( 16\lambda /3\right) ^{2}.  \label{1D}
\end{equation}%
In this context, $N\equiv \left( 3/32\lambda \right) \left(
A_{+}^{2}+A_{-}^{2}\right) $ is again proportional to the norm, and $\nu
\equiv (3/32\lambda )(A_{+}^{2}-A_{-}^{2})$ may be considered as a measure
of the asymmetry. The purport of Eq. (\ref{1D}) is that it predicts a
critical value of $N$ at which the asymmetric solutions emerge, $N_{\mathrm{%
cr}}=1$. As shown in Fig. \ref{fig2}, a principal difference of the SSB for
the CW states from its counterpart for the solitons (see Fig. \ref{fig1}) is
that the present bifurcation is a \textit{supercritical} one.

It is relevant to mention that the SSB for the CW solutions of the
dual-core-fiber model is also supercritical \cite{Snyder}, on the contrary
to the subcritical bifurcation for solitons in the same model \cite{Sydney}
(the bifurcation of the CW solutions may become subcritical if the
nonlinearity is saturable, rather than cubic \cite{Snyder}). Finally, it is
also necessary to mention that all the CW states, considered as quasi-1D
solutions of the 2D model with the self-attracting nonlinearity, are
unstable against modulational perturbations (while solitons may be
completely stable in long simulations, see below).

\section{Numerical results}

To verify the above predictions, we solved Eq.~(\ref{Phi})
numerically, using the imaginary-time relaxation method with a
fourth-order Runge--Kutta algorithm. The accuracy of the numerical
code was tested by varying computational parameters, namely the
mesh density, domain size, and time step. These parameters were
then fixed at values for which further increase of the accuracy
would not lead to a visible change in the final results
\cite{note}. This procedure was applied every time when the
physical parameters $L,D,N$ were varied.

The stability of solitons produced by this method was then tested by direct
integrations of perturbed states in real time. The perturbation was
introduced, multiplying the wave function by a symmetry-breaking factor, $%
\Psi \rightarrow \Psi (1+\alpha y/L)$. Perturbations with $\alpha
=0.05$, which are actually large, were not able to destroy
solitons that were identified as stable ones. On the contrary,
much small perturbations (for instance, with $\alpha =0.002$) were
sufficient to demonstrate instability of those solutions which are
unstable, after propagation time $t=100$.

In this work, we did not attempt to identify stability regions for the
solitons by computing their eigenvalues in terms of linearized equations for
small perturbations, so, in this sense, the stability borders are not
completely rigorous ones. Nevertheless, the distinction between unstable and
stable solitons revealed by the simulations is very clear, and, on the other
hand, the identification of the stability by dint of direct simulations
corresponds to experimental conditions, where solitons are subject to
various perturbations of a finite size.

\begin{figure}[tbp]
\includegraphics[width=8.5cm]{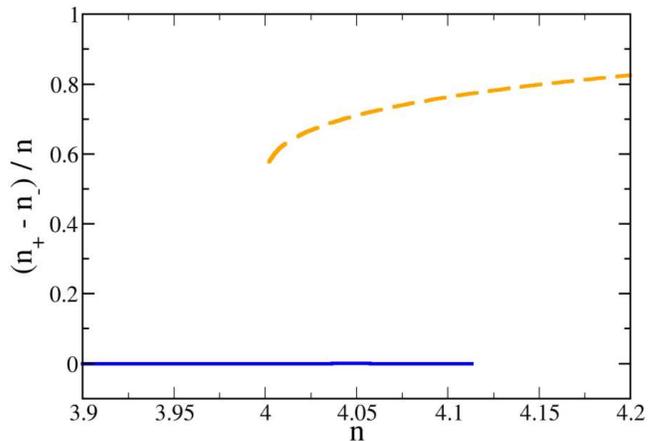}
\caption{(Color online). Imbalance in the norm between half-planes $x>0$ and
$x<0$, defined as per Eq. (\protect\ref{NNN}) for numerically found
stationary soliton solutions, versus the total norm. The continuous and
dashed lined lines show, respectively, numerically found \emph{stable}
symmetric and asymmetric solutions (unstable solutions were not generated by
the numerical procedure). Parameters are: $L=1$, $D=1$, and $U_{0}=1$.}
\label{fig3}
\end{figure}

\begin{figure}[tbp]
\includegraphics[width=8.5cm]{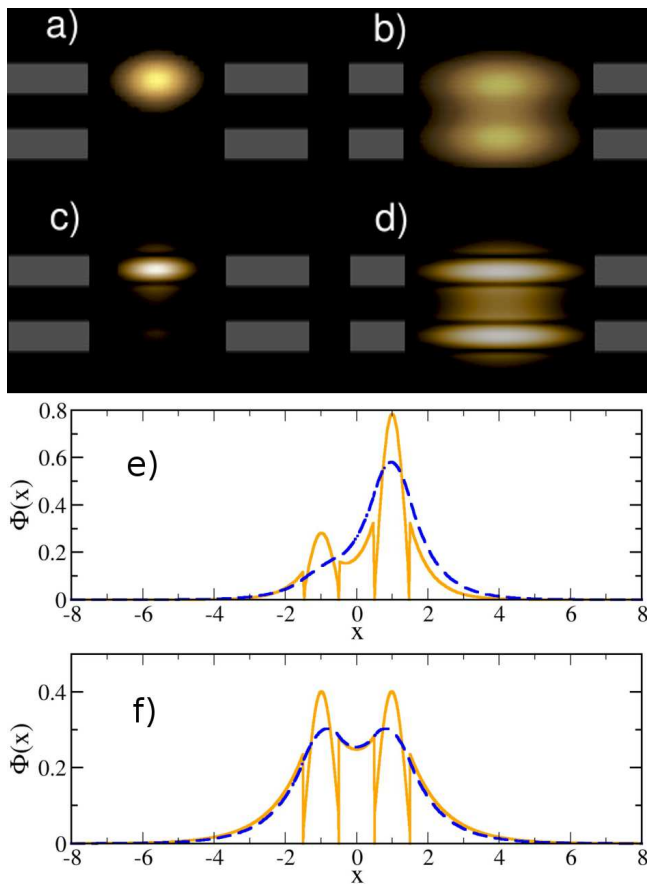}
\caption{(Color online). Examples of numerically found asymmetric (a) and
symmetric (b) soliton solutions of Eq.~(\protect\ref{Phi}), for $n=4.05$.
Other parameters are the same as in Fig.~\protect\ref{fig3}. The gray
shading indicates the position of the two potential wells. Panels (c) and
(d) show the corresponding solutions as predicted by the variational
approximation. In panels (e) and (f), the cross-sections of the solutions
along $y=0$ are compared: The dashed and solid curves correspond to the
numerical and variational solutions, respectively.}
\label{fig4}
\end{figure}

\begin{figure}[tbp]
\includegraphics[width=8.5cm]{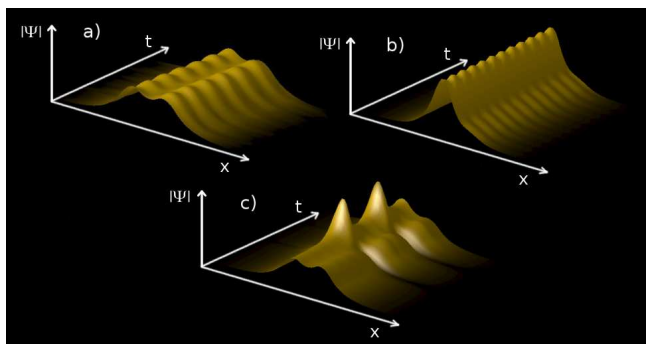}
\caption{Typical examples of the evolution of perturbed stationary states
produced by numerical simulations of Eq.~(\protect\ref{eq1}). (a): A stable
symmetric state for $n=3.9$. (b) and (c): Stable asymmetric and unstable
symmetric states for $n=4.15$. Other parameters are as in Fig.~\protect\ref%
{fig3}. The evolution time is $t=300$.}
\label{figtez4}
\end{figure}

\begin{figure}[tbp]
\includegraphics[width=8.5cm]{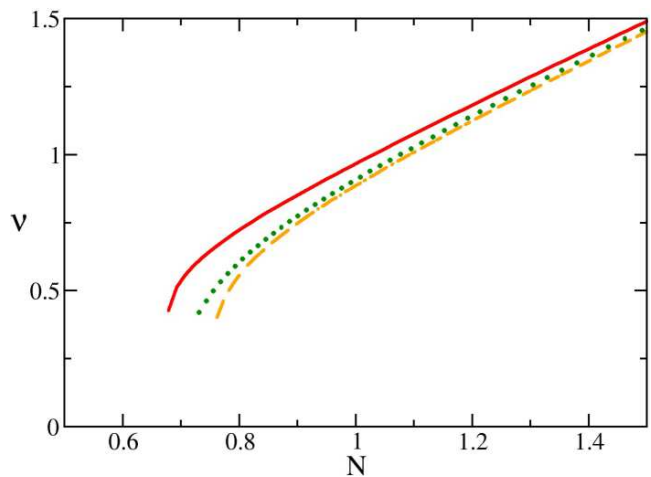}
\caption{(Color online). Comparison of the variational prediction for the
asymmetric solitons (solid line) with respective numerical results (dotted
and dashed lines). The dashed curve corresponds to the same parameters as in
Fig.~\protect\ref{fig3}, while the dotted one is generated by the numerical
solution of Eq. (\protect\ref{Phi}) with the distance between the wells
increased to $L=2$.}
\label{fig5}
\end{figure}

The numerical results are summarized in Figs.~\ref{fig3}-\ref{fig5}. Figure~%
\ref{fig3} displays a typical dependence of the global asymmetry coefficient
for the numerically found soliton solutions, defined as%
\begin{equation}
\frac{n_{+}-n_{-}}{n}\equiv \frac{\int_{-\infty }^{+\infty }dy\left[
\int_{0}^{+\infty }\Phi ^{2}dx-\int_{-\infty }^{0}\Phi ^{2}dx\right] }{%
\int_{-\infty }^{+\infty }\Phi ^{2}dxdy},  \label{NNN}
\end{equation}%
on the total norm, $n$. The way the figure was generated (through direct
simulations converging to stationary states) made is possible to display
only stable branches of the solutions, both symmetric and asymmetric ones.
Although the unstable branches are missing, there is little doubt that the
full SSB diagram corresponding to Fig. \ref{fig3} is of the generic
subcritical type. In particular, a bistability (hysteretic) region, where
symmetric and asymmetric solitons coexist and are simultaneously stable, is
obvious in the figure. Thus, the picture suggested by the numerical results
is fully consistent with the predictions of the VA shown in Fig. \ref{fig1}.

An example of coexisting symmetric and asymmetric solitons is shown in Fig.~%
\ref{fig4}, for the value of the norm $n=4.05$. As both solitons contain
equal total numbers of atoms, the asymmetric one has smaller width in both $x
$ and $y$ directions, as its shape provides for stronger effective
self-attraction. On the the other hand, the shape of the symmetric soliton
features a fair amount of tunneling between the channels. Figures~\ref{fig4}
a) and b) were generated by direct numerical simulations, and Figs.~\ref%
{fig4} c) and d) were obtained from the VA. It is observed that
the agreement between the numerical and variational results is
quite good. For a more detailed comparison, we display
cross-sections of both solitons in panels e) and f).

Figure \ref{figtez4} illustrates the stability of various solitons in
numerical simulations. Below the bifurcation point (for $n=3.9$), we present
stable evolution of the symmetric state, and above the bifurcation we
demonstrate a stable asymmetric state for $n=4.15$. Also, for norm $n=4.15$,
we display the evolution of the unstable symmetric state. It is worthy to
note that this unstable state does not simply relax to the stable one, but
rather performs persistent oscillations between symmetric and asymmetric
shapes.

In Fig.~\ref{fig5}, we present comparison of the analytical results,
obtained by means of the VA for the family of asymmetric solitons, with
numerical findings. The solid line is the $\nu (N)$ dependence corresponding
to the stable upper branch of the plot in Fig.~\ref{fig1}, whereas the other
two lines show two sets of the corresponding numerical results, generated as
said in the caption to Fig. \ref{fig5}. Generally, the solid line
(variational solution) is shifted towards smaller values of $N$.

\section{Conclusions}

In this work, we have introduced a physical model that gives rise to the
first example of the SSB (spontaneous symmetry breaking) of 2D solitons in a
dual-core system; previously, this effect was studied in detail, but solely
in 1D settings. Our model is based on the 2D nonlinear Schr\"{o}dinger
equation with the self-focusing cubic nonlinearity and a quasi-1D
double-channel potential. The model applies to the description of spatial
optical solitons in a bulk medium with two waveguiding slabs embedded into
it, or spatiotemporal solitons in a planar waveguide into which two guiding
channels were inserted. The same model may also be interpreted as the
Gross-Pitaevskii equation for a pancake-shaped Bose-Einstein condensate
trapped around two attractive parallel light sheets. By means of the VA
(variational approximation), we have predicted the SSB bifurcation for the
2D solitons supported by the double channel. The bifurcation was predicted
to be subcritical (unlike its counterpart in the continuous-wave 1D model).
The predictions of the VA are well corroborated by numerical solutions.

\section{Acknowledgements}

M.M. acknowledges support from the Foundation for Polish Science and the
Polish Ministry of Science and Education under grant N202 014 31/0567. M.T.
was supported by the Polish Ministry of Scientific Research and Information
Technology under grant PBZ MIN-008/P03/2003. The work of B.A.M. was
partially supported by the Israel Science Foundation through
Excellence-Center grant No. 8006/03, and by German-Israel Foundation through
grant No. I-884-149.7/2005.


\begin{thebibliography}{99}
\bibitem{note}
 Most of the numerical simulations were performed
using the following set of parameters: time step $\Delta t=0.01$,
mesh density $\Delta x=\Delta y=0.1$, and domain size
$x_{\mathrm{max}}=y_{\mathrm{max}}=8$.
\end{thebibliography}
\end{document}